\documentclass[a4paper]{jpconf}
\usepackage{graphicx}

\usepackage{epsfig,amsfonts,amsthm}
\usepackage{amsmath,amssymb}
\usepackage[normalem]{ulem}
\usepackage{color}
\usepackage{array}
\newcommand{\be}{\begin{equation}}
\newcommand{\ee}{\end{equation}}
\newcommand{\bea}{\begin{eqnarray}}
\newcommand{\eea}{\end{eqnarray}}

\renewcommand{\Re}{\mathrm{Re }}
\renewcommand{\Im}{\mathrm{Im }}

\def\lsim{\mathrel{\rlap{\lower4pt\hbox{\hskip1pt$\sim$}}
    \raise1pt\hbox{$<$}}}         
\def\gsim{\mathrel{\rlap{\lower4pt\hbox{\hskip1pt$\sim$}}
    \raise1pt\hbox{$>$}}}         
    
\usepackage{amsmath}
\usepackage{amsfonts}
\usepackage{amssymb}
\usepackage{subfig}
\usepackage{wrapfig}
\usepackage{graphicx}

\def\beq{\begin{equation}}
\def\eeq{\end{equation}}
\def\bea{\begin{eqnarray}}
\def\eea{\end{eqnarray}}

\def\<{\left\langle}
\def\>{\right\rangle}

\newcommand{\bt}{\begin{tabular}}
\newcommand{\et}{\end{tabular}}

\usepackage{slashed}
\usepackage{amssymb} 
\usepackage{float}

\usepackage[justification=centering]{caption}

\usepackage{tikz}
\usetikzlibrary{decorations.pathmorphing,decorations.markings}

\usepackage{graphicx}

\tikzset{
photon/.style={decorate, decoration={snake,amplitude=2pt, segment length=5pt}, draw=black},
particle/.style={draw=black, postaction={decorate}, decoration={markings,mark=at position .5 with {\arrow[draw=black]{>}}}},
antiparticle/.style={draw=black, postaction={decorate}, decoration={markings,mark=at position .5 with {\arrow[draw=black]{>}}}},
gluon/.style={decorate, draw=black, decoration={coil,amplitude=4pt, segment length=5pt}}
goldstone/.style={draw=green,postaction={decorate},decoration={markings,mark=at position .5 with {\arrow[draw=blue]{>}}}}
}

\begin{document}
\title{Flavon-induced lepton flavour violation}

\author{Venus Keus}

\address{Department of Physics and Helsinki Institute of Physics, Gustaf H{\"a}llstr{\"o}min katu 2, FIN-00014 University of Helsinki, Finland}

\ead{venus.keus@helsinki.fi}

\begin{abstract}
ATLAS and CMS have observed a flavor violating decay of the Higgs to muon and tau. The fact that flavour violating couplings of the Higgs boson are exactly zero in the Standard Model suggests the mixing of the Higgs with another scalar with flavour violating couplings. We use the flavon field from the Froggatt-Nielsen mechanism, responsible for generating the lepton Yukawa matrices, for this purpose. 
The parameter space is constrained from experimental bounds on charged lepton flavor violation in other processes, however, we show that a substantial region of parameter space survives these bounds while producing a large enough $Br(h \rightarrow  \mu \tau)$.
\end{abstract}

\section{Introduction}
Fermion Yukawa couplings in the Standard Model (SM) are free parameters and there is no explanation for the hierarchy among the fermion masses within the SM. 
A popular Beyond Standard Model (BSM) framework that naturally generates the SM fermion Yukawa couplings was suggested by Froggatt and Nielsen \cite{Froggatt:1978nt}. 

Lepton Flavour Violating (LFV) couplings of the Higgs boson are zero in the SM, however, ATLAS and CMS experiments have reported on a LFV decay of the observed boson \cite{Khachatryan:2015kon, Aad:2015gha}\footnote{The process $h\to l_i l_j$ is assumed to be the sum of $h\to l_i^{+}l_j^{-}$ and $h\to l_i^{-}l_j^{+}$.} at the LHC with the combined branching ratio
\be 
Br(h\to \mu \tau) = 0.82^{+0.33}_{-0.32} .
\label{h-tau-mu}
\ee
Recent results from the LHC \cite{Aad:2016blu} has reduced this value to $0.55^{+0.28}_{-0.28}$.

In a recent paper \cite{Huitu:2016pwk}, we aimed at explaining LFV decays of the Higgs by employing the Froggatt-Nielsen mechanism. This mechanism requires the addition of a scalar field, \emph{flavon}, singlet under the SM gauge group, charged under an extra $U(1)$-symmetry which breaks spontaneously due to the flavon field acquiring a Vacuum Expectation Value (VEV) which then generates the SM Yukawa matrices through higher order operators. 
The Higgs boson then acquires LFV couplings through mixing with the flavon.

The resulting mass eigenstates could then mediate LFV processes which are constrained by experimental data and exclude regions of the parameter space. We show that it is possible to acquire a large enough $Br(h\to\mu\tau)$ in the surviving regions of the parameter space of the model.

\section{The Froggatt-Nielsen framework}
\label{Froggatt-Nielsen-review}
In the Froggatt-Nielsen mechanism the SM Yukawa interaction are generated through higher order operators which are
obtained by integrating out heavy states at some scale $\Lambda$. Such operators  which are invariant under a $U(1)$ symmetry are of the form 
\begin{equation}
c_{ij} ~ { \Phi^{n_{ij}} \over \Lambda^{n_{ij} } }   \bar f_{L,i} f_{R,j} ~H  + {\rm h.c.} ~,
\label{operator}
\end{equation} 
where $c_{ij}$ are order one coefficients, $f_{L,R}$ are SM fermions and $\Phi$ is the flavon which acquires a VEV,
\begin{equation}
\Phi = {1\over \sqrt{2}} (v_\phi + \phi)
\end{equation}
with $\phi=\Re\phi+ i \Im\phi$. The Yukawa couplings are then given by 
\begin{equation}
 Y_{ij}= c_{ij} ~ \left( { v_\phi \over \sqrt{2} \Lambda }\right)^{n_{ij} } \equiv c_{ij}~ \epsilon^{n_{ij} } \;,
 \label{Y}
\end{equation}
where $\epsilon$ is a small parameter. 
The $U(1)$ symmetry requires 
\be 
 n_{ij} =-\frac{1}{q_{\phi}}(q_{\bar L,i}+q_{R,j}+q_{h}) \;,
\label{flavon power}
\ee 
where $q_i$ are the charges identified in Table \ref{U(1) charges}. 

\begin{table}
\begin{center}
\begin{tabular}{lllll}
\br
Particle   &  $ f_{L,i}^c$ & $f_{R,i}$ & $H$ & $\Phi$\\
\mr
$U(1)$ charge & $q_{\bar L,i}$ & $q_{R,i}$ & $q_{h}$ & $q_{\phi}$ \\
\br
\end{tabular}
\end{center}
\caption{The $U(1)$ charges of SM fermions $f_{R,L}$, SM Higgs field $H$ and the flavon $\Phi$.}
\label{U(1) charges}
\end{table}

In principle, one can introduce several flavons each responsible for describing the hierarchy of the masses in the quark, charged and neutral lepton sectors. Here, we focus on the leptonic sector and allow for only the flavon and the leptons to transform under the $U(1)$ symmetry. As a result, after the Higgs doublet acquires a VEV, 
\be 
H = \left(\begin{array}{c}
0\\
\frac{v+h}{\sqrt{2}}
\end{array}\right) \qquad 
 \label{field-definition}
\ee 
the effective Lagrangian takes the form 
\bea
\label{L-eff}
\mathcal{L}_{eff}&\supset &
 \frac{v}{\sqrt{2}} ~ Y_{ij} ~ \left(      1+ {h\over v} +  n_{ij}~{\phi \over v_\phi} \right)   \bar{l}'_{L,i}l'_{R,j} \;.   
 \eea

Here, $l'$ stands for the lepton fields in the weak basis and $l$ for the mass eigenstate basis. 
Rotating the lepton fields from gauge eigenstates to the mass eigenstates requires diagonalising the lepton Yukawa matrix which is done using two unitary matrices:
\be 
Y_{\textrm{diag}}=U_{L}Y~U_{R}^{\dagger}.
\ee
The effective Lagrangian then becomes
\bea
\mathcal{L}_{eff}&\supset & 
\frac{v}{\sqrt{2}} \bar{l}_{L} \; Y_{\textrm{diag}} \;l_{R} +
 {h\over \sqrt{2}} ~ \bar{l}_{L} \; Y_{\textrm{diag}} \; l_{R}
 + {\phi \over \sqrt{2}} {v\over v_\phi} ~\bar{l}_{L} \;  \kappa  \; l_{R}
   + {\rm h.c.},  
   \label{LFVinteraction}
\eea
where the flavon 
vertex involves the matrix 
\begin{equation}
\kappa = U_{L} \; (Y_{ij} n_{ij}) \; U_{R}^{\dagger}~,
\end{equation}
with $n_{ij}=a_i +b_j$. Setting 
$q_\Phi=-1$ and $q_h=0$
throughout this paper, we have 
\be
\kappa_{ij}= y_{j}\sum_{k=1}^{3}q_{\bar L,k}(U_{L})_{ik}(U_{L})^{\ast}_{jk}
+y_{i}\sum_{k=1}^{3}q_{R,k}(U_{R})_{ik}(U_{R})^{\ast}_{jk} ~,
\label{kappa}
\ee
where $y_{i}$ are the Yukawa matrix eigenvalues. This is the source of lepton flavor violation in our model. Upon the Higgs--flavon mixing, such flavor changing couplings 
also appear in the interactions of the physical Higgs--like boson.

\subsection{The scalar potential}
The $U(1)$-symmetric scalar potential is of the form 
\be 
V(H,\Phi)=-\mu_h^2(H^\dagger H) + \lambda_h(H^\dagger H)^2 -\mu_\phi^2(\Phi^\dagger \Phi) + \lambda_\phi(\Phi^\dagger \Phi)^2 + \lambda_{h\phi} (H^\dagger H)(\Phi^\dagger \Phi)~.
\ee
The minimisation conditions lead to the following mass eigenstates $H_1$ and $H_2$ which are rotated by the following matrix
\be 
\left(\begin{array}{c}
H_1\\
H_2
\end{array}\right) \equiv 
\left(\begin{array}{cc}
\cos\theta & \sin\theta\\
-\sin\theta & \cos\theta 
\end{array}\right) \left(\begin{array}{c}
h\\
\Re\phi
\end{array}\right) 
\ee
where the mixing angle is defined as
\be 
\tan 2\theta = \frac{\lambda_{h\phi} v_h v_\phi}{\lambda_{h} v_h^2 -\lambda_\phi v_\phi^2}
\ee
and is constrained by theoretical and experimental data, reviewed and summarised for Higgs-portal models in \cite{Falkowski:2015iwa} which are taken into account in our calculations.
The masses of $H_1$ and $H_2$ are as follows
\bea 
m^2_{H_{1,2}}&=& \lambda_{h} v_h^2 +\lambda_\phi v_\phi^2  \pm \sqrt{\biggl(\lambda_{h} v_h^2 -\lambda_\phi v_\phi^2  \biggr)^2 + \lambda_{h\phi}^2 v_h^2 v_\phi^2} 
\eea
where we take the lighter boson, $H_1$, to be the SM-like Higgs boson, the observed scalar at the LHC with mass $125$ GeV.

In the case of a global $U(1)$ symmetry which is broken by the flavon VEV, the model predicts a massless Goldstone boson which is    phenomenologically unacceptable. In the case of a gauged $U(1)$ symmetry, we find that  due to tight constraints on a flavor non-universal $Z^\prime$, the model does not lead to a large enough $Br(h\to\mu\tau)$. Here, we assume a discretized $U(1)\;\rightarrow \;Z_N$ by introducing a soft explicit breaking term ${ \tilde m}^2 \Phi^2 +  {\rm h.c.}$ which induces $m_{{\rm Im} \phi} \sim \tilde m$. 
Note that our approach is one of an effective field theory and while tree level processes are well under control, the loop contributions are only indicative in nature and depend on the details of the UV completion.

\section{Bounds on lepton flavor violating processes}
\label{CLFV-bounds}

\begin{table}
\begin{center}
\begin{tabular}{lll}
\br 
 & Observable & Present limit \\[1mm]
\mr
1 & BR$(\mu\to eee)$  & $1.0\times 10^{-12}$ \cite{Bellgardt:1987du} 
\\[2mm]
2 & BR$(\tau\to eee)$  & $3.0\times 10^{-8}$ \cite{Amhis:2012bh} 
\\[2mm]
3 & BR$(\tau\to \mu\mu\mu)$  & $2.0\times 10^{-8} $ \cite{Amhis:2012bh}
\\[2mm]
4 & BR$(\tau^{-}\to\mu^{-}e^{+}e^{-})$  & $1.7\times 10^{-8}$ \cite{Hayasaka:2010np}
\\[2mm]
5 & BR$(\tau^{-}\to e^{-}\mu^{+}\mu^{-})$ & $2.7\times 10^{-8}$ \cite{Hayasaka:2010np}
\\[2mm]
6 & BR$(\tau^{-}\to e^{+}\mu^{-}\mu^{-})$ & $1.7\times 10^{-8}$ \cite{Hayasaka:2010np}
\\[2mm]
7 & BR$(\tau^{-}\to \mu^{+}e^{-}e^{-}$) & $1.5\times 10^{-8}$ \cite{Hayasaka:2010np}
\\[2mm]
8 & BR$(\mu\to e\gamma)$ & $5.7\times 10^{-13}$ \cite{Adam:2013mnn} 
\\[2mm]
9 & BR$(\tau\to \mu\gamma)$  & $4.4\times 10^{-8}$ \cite{Amhis:2012bh}
\\[2mm]
10 & BR$(\tau\to e\gamma)$ & $3.3\times 10^{-8}$ \cite{Amhis:2012bh}
\\[2mm]
11 & CR$(\mu$-$e, Au)$ & $7.0\times 10^{-13}$ \cite{Bertl:2006up}
\\[1mm]
\br 
\end{tabular}
\caption{Current experimental bounds on the branching ratios of three--body LFV decays, magnetic transitions   and the conversion rate of $\mu \to e$.}
\label{experimental-bounds}
\end{center}
\end{table}

The flavon interaction in equation (\ref{LFVinteraction}) induces LFV processes mediated by the three scalar mass eigenstates
which are strongly constrained by experiment. We impose the current bounds from 
 the three--body decay  $l_{i}\to l_{j}l_{k}l_{l}$, magnetic transition $l_{i}\to l_{j}\gamma$ and $\mu \to e$ conversion processes presented in Table \ref{experimental-bounds} which put limits on the flavon couplings parametrized by
\begin{equation}
\tilde \kappa_{ij} = {1 \over \sqrt{2}} {v \over v_\phi} \kappa_{ij} \;,
\end{equation}
where the flavon--lepton coupling is  $ \tilde \kappa_{ij} \; \bar{l}_{L,i}  \; l_{R,j} \phi+ $ h.c.

We assign charges listed in Table \ref{charges-table1} to produce the following 
Yukawa texture and $\tilde \kappa$ matrix
\be
Y = \left(
\begin{array}{ccc}
 3.4 ~\epsilon^6 & -0.6 ~\epsilon^6 & 3.5~ \epsilon^7 \\
 5.4 ~\epsilon^4 & 6.1 ~\epsilon^4  & -3.1 ~\epsilon^5 \\
 0.5 ~\epsilon^2  & 0.5 ~\epsilon^2    & 7.3 ~\epsilon^3 \\
\end{array}
\right), \quad 
\tilde \kappa = \frac{v}{v_\phi}\left(
\begin{array}{ccc}
1\times 10^{-5}  & -1\times 10^{-6}  & -3\times 10^{-6}  \\
-2\times 10^{-5}  & 2\times 10^{-3} & 6\times 10^{-4}  \\
3\times 10^{-4} & -4\times 10^{-3} & 2\times 10^{-2}
\nonumber
\end{array}
\right),
\label{exact-texture}
\ee
which reproduces the correct lepton masses for $\epsilon=0.1$ and the shown order one coefficients (their precise values are given in \cite{Huitu:2016pwk}). 

\begin{table}
\begin{center}
\begin{tabular}{lllllllll}
\br
Particle   &  $e_L^c$ & $e_R$ & $\mu_L^c$ & $\mu_R$ & $\tau_L^c$ &$\tau_R$ & $H$ & $\phi$
\\
\mr
Charge & 6 & 0 & 4 & 0 & 2 & 1 & 0 & -1\\
\br
\end{tabular}
\end{center}
\vspace{-3mm}
\caption{U(1)/Z$_N$ charge assignment. }
\label{charges-table1}
\end{table}

\subsection{Negligible Higgs--flavon mixing }
It is instructive to consider a scenario where the Higgs-flavon mixing is close to zero in which case only Re$\phi$ and Im$\phi$ mediate the LFV processes. To make the discussion clearer, we also decouple the Im$\phi$ in this sub-section.

\begin{minipage}{\linewidth}
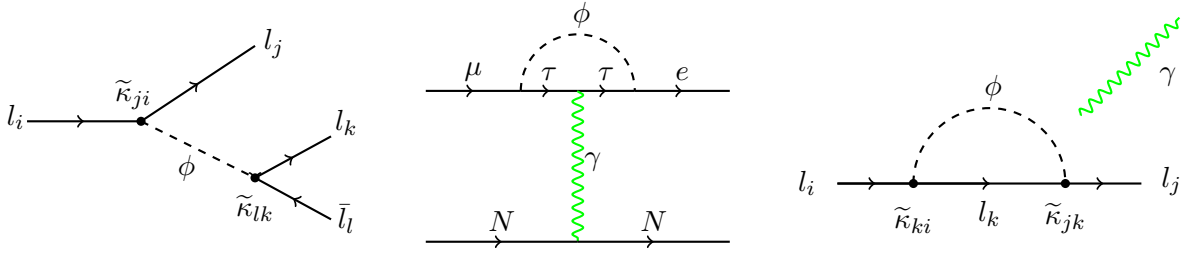
\begin{figure}[H]
\begin{tikzpicture}[thick,scale=1.0]
\fill[black] (1.5,0) circle (0.06cm);
\draw (1.5,0) -- node[black,above,xshift=-0.1cm,yshift=0.0cm] {$\widetilde{\kappa}_{j i}$} (1.5,0.03);
\draw[particle] (0,0) -- node[black,above,xshift=-0.9cm,yshift=-0.25cm] {$l_{i}$} (1.5,0);
\draw[particle] (1.5,0) -- node[black,above,xshift=1.0cm,yshift=0.2cm] {$l_{j}$} (3,1.0);
\draw[particle] (3,-0.75) -- node[black,above,yshift=0.15cm,xshift=0.7cm] {$l_{k}$} (4,-0.2);
\draw[particle] (4,-1.3) -- node[black,above,yshift=-0.6cm,xshift=0.7cm] {$\bar{l}_{l}$} (3,-0.75);
\draw[dashed] (1.5,0) -- node[black,above,yshift=-0.55cm,xshift=-0.15cm] {$\phi$} (3,-0.75);
\fill[black] (3,-0.75) circle (0.06cm);
\draw[particle] (3,-0.75) -- node[black,above,yshift=-0.7cm,xshift=0.0cm] {$\widetilde{\kappa}_{l k}$} (3,-0.78);
\end{tikzpicture}
\hspace{5mm}
\begin{tikzpicture}[thick,scale=1.0]
\draw[particle] (0,0) -- node[black,above,sloped,yshift=0.0cm,xshift=0.0cm] {$\mu$} (1.25,0);
\draw[particle] (1.25,0) -- node[black,above,sloped,yshift=0.0cm,xshift=0.0cm] {$\tau$} (2,0);
\draw[particle] (2,0) -- node[black,above,sloped] {$\tau$} (2.75,0);
\draw[particle] (2.75,0) -- node[black,above,sloped] {$e$} (4,0);
\draw[decorate,decoration={snake,amplitude=2pt,segment length=5pt},green] (2,0) -- node[black,above,yshift=-0.2cm,xshift=0.2cm] {$\gamma$} (2,-2);
\draw[particle] (0,-2) -- node[black,above,sloped,yshift=0.0cm,xshift=0.0cm] {$N$} (2,-2);
\draw[particle] (2,-2) -- node[black,above,sloped,yshift=0.0cm,xshift=0.0cm] {$N$} (4,-2);
\draw[dashed]  (1.25,0) node[black,above,sloped,yshift=0.7cm,xshift=0.8cm] {$\phi$}  arc (180:0:0.75cm) ;
\end{tikzpicture}
\hspace{5mm}
\begin{tikzpicture}[thick,scale=1.0]
\fill[black] (1,0) circle (0.06cm);
\draw (0,0) -- node[black,above,yshift=-0.8cm,xshift=0.0cm] {$\widetilde{\kappa}_{k i}$} (2,0);
\draw[particle] (0,0) -- node[black,above,sloped,yshift=-0.3cm,xshift=-0.9cm] {$l_{i}$} (1,0);
\draw[particle] (1,0) -- node[black,above,sloped,yshift=-0.7cm,xshift=0.0cm] {$l_{k}$} (3,0);
\draw[particle] (3,0) -- node[black,above,sloped,yshift=-0.3cm,xshift=0.9cm] {$l_{j}$} (4,0);
\draw[decorate,decoration={snake,amplitude=2pt,segment length=5pt},green] (3.2,0.9) -- node[black,above,yshift=-0.4cm,xshift=0.5cm] {$\gamma$} (4.5,2.2);
\draw[dashed]  (1,0) node[black,above,sloped,yshift=0.95cm,xshift=1.05cm] {$\phi$}  arc (180:0:1cm) ;
\fill[black] (3,0) circle (0.06cm);
\draw (3,0)  node[black,above,yshift=-0.8cm,xshift=0.0cm] {$\widetilde{\kappa}_{j k}$} (3,0);
\end{tikzpicture}
\vspace{0.5cm}
\caption{The $l_{i}\to l_{j}l_{k}l_{l}$ (left), $\mu\leftrightarrow e$-conversion (center) and $l_{i}\to l_{j}\gamma$ (right) processes mediated by the flavon  $\phi$.  The decay $l_{i}\to l_{j}l_{k}l_{l}$ also receives important contributions at one loop.}
\label{CLFV-processes-fig} 
\end{figure}    
\end{minipage}

\vspace{2mm}

We have taken into account all processes represented in Fig.~\ref{CLFV-processes-fig} at tree and loop level. Here, we only show  
the details of the $\mu \to e \gamma$ process which puts the strongest bounds on the model.  
Neglecting the light lepton contributions, we find 
\begin{eqnarray}
\Gamma(\mu\to e\gamma)&=&
\frac{\alpha m_{\mu}^{3}m_{\tau}^{2}}{1024\pi^{4}} ~ \frac{|\widetilde{\kappa}_{e\tau}|^2 |\widetilde{\kappa}_{\tau\mu}|^2 + |\widetilde{\kappa}_{\tau e}|^2|\widetilde{\kappa}_{\mu\tau}|^2}{m_{{\rm Re} \phi}^{4}} \left[\frac{3}{2}-\log\left(\frac{ m_{\phi}^{2}}{m_{\tau}^{2}}\right)\right]^{2},
\end{eqnarray}

The invariant amplitude $A_{\mu \to e\gamma}^{L}$ is
\be
\label{invariant amplitude mu}
A_{\mu\to e\gamma}^{L}=-\frac{ie}{32\pi^{2}}\widetilde{\kappa}_{\tau e}^{\ast}\widetilde{\kappa}_{\mu\tau}^{\ast}\left[\frac{3}{2}-\log\left(\frac{m_{\phi}^{2}}{m^{2}_{\tau}}\right)\right]\frac{m_{\tau}}{m_{\phi}^{2}}~
\ee
and the corresponding equation for $A^{R}$ is obtained by replacing $\widetilde{\kappa}_{ij}$ with $\widetilde{\kappa}_{ji}^*$.  

\begin{figure}[h!]
\begin{center}
\includegraphics[scale=0.7]{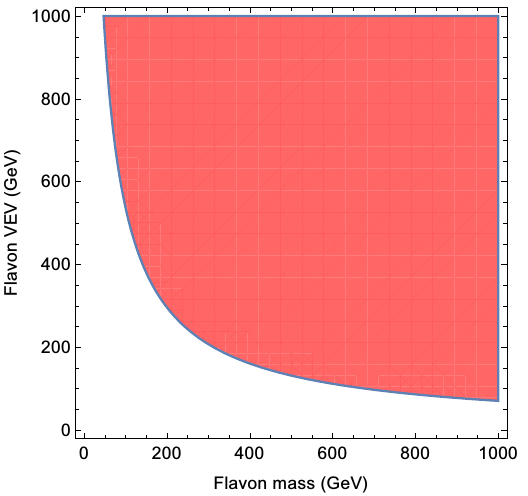}
\end{center}
\vspace{-3mm}
\caption{ Allowed parameter space (shaded) for the texture at hand (Eq.\ref{exact-texture}) with negligible Higgs-flavon mixing. The strongest constraint is imposed by $Br(\mu \rightarrow e \gamma)$.}
\label{Zeromixing}
\end{figure}

Figure \ref{Zeromixing} shows the allowed values of 
$v_\phi$ vs $m_{\phi}$ after imposing the $\mu\to e \gamma$ bounds. 

\vspace{-3mm}

\subsection{Substantial Higgs--flavon mixing }
In the case of non-zero Higgs-flavon mixing, all three scalar  mass eigenstates, $H_1, H_2$ and Im$\phi$, mediate LFV processes. The relevant interaction terms are
 \bea
\mathcal{L} &\supset& 
\left[\cos\theta~\frac{Y^{\rm diag}_{ij}}{\sqrt{2}}+\sin\theta~\widetilde{\kappa}_{ij}\right]
\bar{l}_{i}P_{R}l_{j} \;H_{1}
 +\left[-\sin\theta~\frac{Y^{\rm diag}_{ij}}{\sqrt{2}}+\cos\theta~\widetilde{\kappa}_{ij}\right]\bar{l}_{i}P_{R}l_{j} \; H_{2} 
 \nonumber\\
&+&
i  \widetilde{\kappa}_{ij} \; \bar{l}_{i}P_{R}l_{j}  \; \Im\phi 
+ {\rm h.c.}   
\label{couplings in mixing}
\eea
The couplings of $H_1, H_2$ to quarks are flavor--diagonal and are rescaled compared to the corresponding SM couplings by $\cos\theta$ and  $-\sin\theta$, respectively.

Note that in this case due to the couplings of $H_1$ and $H_2$ to quarks, the $\mu \leftrightarrow e$ conversion process is possible at tree level and the $\mu \to e \gamma$ processes gets significant contributions from 2--loop Barr-Zee diagrams with the top quark and the W in the loop. 

Again, the most limiting constraints comes from the  $\mu \rightarrow e\gamma$ process with the 1-- and 2--loop contributions to the amplitude
\be
A_{\mu\to e\gamma}^{L}=A_{\mu\to e\gamma}^{L}(1-\textrm{loop})+A_{\mu\to e\gamma}^{L}(2-\textrm{loop}) ~.
\ee
At one loop we have
\bea
A_{\mu\to e\gamma}^{L}(1-\textrm{loop})&=&
-\frac{ie m_{\tau}}{32\pi^{2}}\widetilde{\kappa}^{\ast}_{\tau e}\widetilde{\kappa}^{\ast}_{\mu\tau}
\biggl\{
\frac{\sin^{2}\theta}{m_{H_{1}}^{2}}\left[\frac{3}{2}-\log\left(\frac{m_{H_{1}}^{2}}{m^{2}_{\tau}}\right)\right]  \\
&&~~~~~~~~~~~~~~~
+\frac{\cos^{2}\theta}{m_{H_{2}}^{2}}\left[\frac{3}{2}-\log\left(\frac{m_{H_{2}}^{2}}{m^{2}_{\tau}}\right)\right]
-\frac{1}{m_{\Im\phi}^2}\left[\frac{3}{2}-\log\left(\frac{m_{\Im\phi}^{2}}{m^{2}_{\tau}}\right)\right]
\biggr\}.  \nonumber
\eea
The 2--loop amplitude receives contributions from the top quark and the $W$ boson \cite{Harnik:2012pb},
\be
A_{\mu\to e\gamma}^{L}(2-\textrm{loop})=A^{L}_{t}+A^{L}_{W},
\ee
with  
\be
A^{L}_{t}=-i\frac{e\alpha v G_{F} }{6\sqrt{2}\pi^3} \sin\theta\cos\theta \;\widetilde{\kappa}_{\mu e}^{\ast} \left[f(z_{tH_{1}})-f(z_{tH_{2}})\right]\;,
\ee
and
\bea
A^{L}_{W}&=&i\frac{e\alpha v G_{F}}{16\sqrt{2}\pi^3}\sin\theta\cos\theta \; \widetilde{\kappa}_{\mu e}^{\ast}\\
&&\times\left\{\left[3f(z_{WH_{1}})+5g(z_{WH_{1}})+\frac{3}{4}g(z_{WH_{1}})+\frac{3}{4}h(z_{WH_{1}})+\frac{f(z_{WH_{1}})-g(z_{WH_{1}})}{2z_{WH_{1}}} \right]\right.\nonumber\\
&&-\left.\left[3f(z_{WH_{2}})+5g(z_{WH_{2}})+\frac{3}{4}g(z_{WH_{2}})+\frac{3}{4}h(z_{WH_{2}})+\frac{f(z_{WH_{2}})-g(z_{WH_{2}})}{2z_{WH_{2}}} \right]\right\}.\nonumber
\eea
Here the loop functions are:
\bea
f(z)&=&\frac{1}{2}z\int_{0}^{1}dx\frac{1-2x(1-x)}{x(1-x)-z}\log\left(\frac{x(1-x)}{z}\right),\\
g(z)&=&\frac{1}{2}z\int_{0}^{1}dx\frac{1}{x(1-x)-z}\log\left(\frac{x(1-x)}{z}\right),\\
h(z)&=&z^{2}\frac{\partial}{\partial z}\left(\frac{g(z)}{z}\right)=\frac{z}{2}\int_{0}^{1}\frac{dx}{z-x(1-x)}\left[1+\frac{z}{z-x(1-x)}\log\left(\frac{x(1-x)}{z}\right)\right].
\eea
The arguments of these functions are defined by $z_{tH_{i}}=m_{t}^{2}/m^{2}_{H_{i}}$ and $z_{WH_{i}}=m_{W}^{2}/m^{2}_{H_{i}}$, with $i=1,2$. The $A_{\mu\to e\gamma}^{R}(2-\textrm{loop})$ amplitude is obtained by replacing  $\widetilde{\kappa}^{\ast}_{ji}$ with  $\widetilde{\kappa}_{ij}$. 
The $\mu\to e\gamma$  decay width is then calculated to be
\be
\Gamma(\mu\to e\gamma)=\frac{m_{\mu}^{3}}{4\pi}\left(\lvert A^L\rvert^{2}+\lvert A^R\rvert^{2}\right)\;.
\ee

\begin{figure}[ht!]
\begin{center}
\includegraphics[scale=0.52]{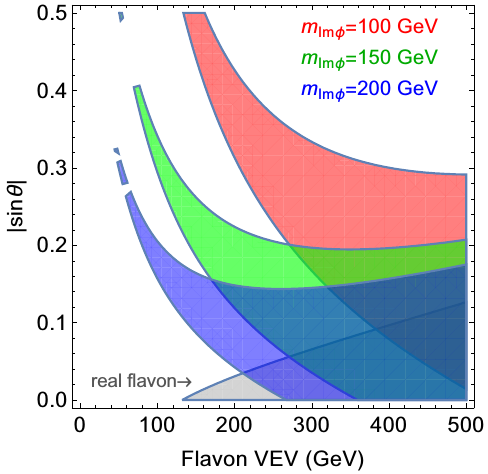}
\includegraphics[scale=0.68]{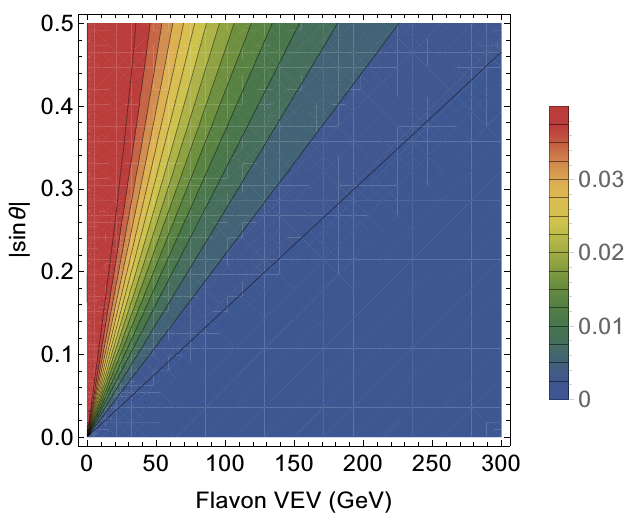}
\end{center}
\caption{Left:   parameter space allowed by the LFV constraints  for $m_{\Im\phi}=100,\;150, \;200$ GeV. We have set $m_{H_2}= 500$ GeV. (The discontinuities appear for technical reasons.) Right: BR$_{\rm eff} (H_{1}\to\mu\tau)$ as a function of $v_\phi$ and $\vert \sin\theta\vert$.
}
\label{1-LHC}
\end{figure}

In Fig.~\ref{1-LHC}  (left) we show the surviving region in the $(v_\phi, \sin\theta)$ plane after imposing these LFV bounds for and exemplary value of $m_{H_2}=500$ GeV.

\vspace{-3mm}   
   
We see that for $\vert \sin\theta\vert \simeq 0.3$ and $v_\phi \sim 100$ GeV the model survives the LFV bounds for a range of $m_{\Im\phi}$ around 150-200 GeV. Note that direct collider constraints on Im$\phi$ are very loose due to its small couplings to leptons. Fig.~\ref{1-LHC} (right) then shows that one expects a substantial decay rate $H_{1}\to\mu\tau$ for $v_\phi \sim 100$ GeV.

It is important to note that such a scenario also leads to the  
enhancement of the lepton diagonal couplings of Higgs which are constrained by the LHC data. We choose $\sin\theta<0$ which leads to some cancellations in $H_1 \rightarrow l_i l_i$ for our Yukawa texture. 
We have
  \bea
\Gamma(H_{1}\to\mu\tau)&=&
\frac{m_{H_{1}}}{8\pi}\sin^{2}\theta \; \left(\vert \widetilde{\kappa}_{\mu\tau}\vert^2+\vert \widetilde{\kappa}_{\tau\mu}\vert^2\right) \\
\Gamma(H_{1}\to\tau\tau) &=&  
 \frac{m_{H_{1}}}{8\pi}\left[\cos\theta\; \frac{Y_{\tau}^{\rm diag}}{\sqrt{2}}+\sin\theta \; \widetilde{\kappa}_{\tau\tau}\right]^{2} \;,
\eea
and analogously for $H_1 \rightarrow \mu\mu$. In our convention, a negative $\theta$ reduces  $\Gamma(H_{1}\to\tau\tau)$ without affecting the LFV rates.

\begin{figure}[ht!]
\begin{center}
\includegraphics[scale=0.72]{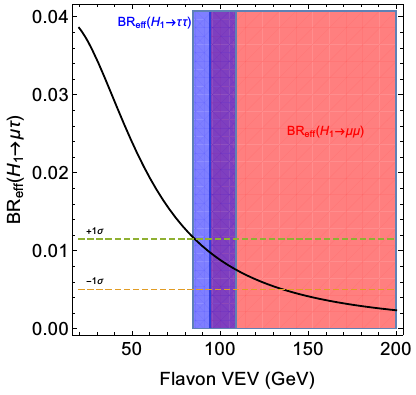}
\end{center}
\caption{  Br$_{\rm eff} (H_1 \to \mu \tau) $ vs $v_\phi$ (black curve)  for  $\sin\theta=-0.3$. The red region is allowed by  ${\rm Br_{eff}  }(H_1 \rightarrow \mu\mu)$ at 95\% CL, the blue region is allowed  by ${\rm Br_{eff}  }(H_1 \rightarrow \tau\tau)$, while their overlap (purple) is consistent with both. The dashed lines show the $\pm 1 \sigma$ limits on the observed Br$_{\rm eff} (H_1 \to \mu \tau)$.
  }
\label{2-LHC}
\end{figure}

Lastlly, to incorporate the difference between the $H_1$ and $h$ production cross sections, we introduce the effective branching ratio 
${\rm Br_{eff} }(H_1 \rightarrow l_i l_j)$ through 
 \begin{equation}
 \sigma(H_1)\; {\rm Br }(H_1 \rightarrow l_i l_j) = \sigma(h)\;  { \Gamma(H_1 \rightarrow l_i l_j)  \over  \Gamma^{\rm total}_{\rm SM} (h) } \equiv \sigma(h)\; {\rm Br_{eff} }(H_1 \rightarrow l_i l_j) \;,
 \end{equation}
and apply the LHC result in equation (\ref{h-tau-mu}) to this effective branching ratio.
We show our results in Fig. \ref{2-LHC} where all of the constraints are satisfied and the observed BR$_{\rm eff} (H_{1}\to\mu\tau)$  for $\sin \theta=-0.3$ is presented.  

\vspace{-3mm}

\section{Conclusion}
\label{conclusion}
 
We have employed a lepton--specific Froggatt--Nielsen framework which naturally generates the lepton Yukawa couplings and leads to lepton flavor violation at the observable level. The flavon predicted by this mechanism mixes with the SM Higgs boson and introduces a $h\to \mu\tau$ decay channel with the branching ratio at the percent level with a flavon VEV at the electroweak scale. We have found this scenario to be consistent with the LFV  and Higgs data constraints. 

\ack 
VK's research is partially supported by the Academy of Finland project ``The Higgs Boson and the Cosmos'', project no. 267842. She also acknowledges the H2020-MSCA-RISE-2014 grant no. 645722 (NonMinimalHiggs) and would like to thank the organizers of DISCRETE 2016 for the kind hospitality and the lively scientific atmosphere provided by the conference.

\end{document}